\newcommand{\CLASSINPUTinnersidemargin}{13mm}

\newcommand{\CLASSINPUToutersidemargin}{13mm}
\documentclass[conference]{IEEEtran}
\usepackage{graphicx}
\usepackage{amsmath}
\usepackage{amssymb}
\usepackage{multirow}
\usepackage{cite}
\usepackage[utf8]{inputenc}
        \usepackage[T1]{fontenc}
\usepackage{amsthm}
\usepackage{algorithm}
\usepackage{algpseudocode}

\usepackage{hyperref}
\usepackage{cleveref}

\begin{document}

\newcommand{\mat}{\mathbf}
\renewcommand{\vec}[1]{\bar{\mat{#1}}}
%
\title{On Energy Cooperation in Energy Harvesting Underlay Cognitive Radio Network}

\author{\IEEEauthorblockN{Kalpant~Pathak and Adrish~Banerjee}
\IEEEauthorblockA{Department of Electrical Engineering, Indian Institute of Technology Kanpur, Uttar Pradesh, 208016\\
Email: \{kalpant, adrish\}@iitk.ac.in}

}

%

\maketitle

\begin{abstract}
In this paper, we consider an energy harvesting cognitive radio network (EH-CRN), where a primary and a secondary user coexist in underlay mode. Both the transmitters have energy harvesting capability and are equipped with finite capacity battery to store the harvested energy. In addition, the secondary user (SU) has an independent energy transfer unit such that it can transfer some portion of it's harvested energy to the primary user (PU). We obtain an optimal transmit power and energy transfer policy for single-slot and a suboptimal policy for multi-slot scenario maximizing the number of bits transmitted by SU under the primary sum-rate constraint in an offline setting. For both cases, the effect of energy cooperation on the system performance is studied and it is observed that energy cooperation results in higher SU throughput.
\end{abstract}

\IEEEpeerreviewmaketitle


\section{Introduction}
\indent Energy harvesting cognitive radio network (EH-CRN) has emerged as a promising solution to the problem of spectrum scarcity while achieving green communication at the same time \cite{survey_1,survey_2,eh_cognitive}. EH-CRNs operating in interweave, overlay and underlay mode, have been extensively considered in the literature \cite{eh_cognitive_1,eh_cognitive_3,eh_cognitive_4,
eh_cognitive_5,eh_cognitive_6,coop_cognitive_shafie,yin_cooperation,
info_ener_coop_cognitive,sanket_pimrc,eh_cognitive_2,rakovic_underlay}.

In interweave mode, the secondary users (SUs) search for vacant spectrum bands in primary channel and access the primary channel opportunistically. In \cite{eh_cognitive_1}, authors have considered energy harvesting interweave cognitive radio network. In the system model considered, primary user (PU) has energy harvesting capability whereas the secondary user (SU) has a reliable power supply. The SU not only accesses the channel in absence of PU but also transmits along with PU with a probability $p$. The authors obtained the optimal probability of access $p$ that maximizes SU's throughput. In \cite{eh_cognitive_3}, authors considered an energy harvesting SU that senses and accesses the primary channel randomly, making use of primary's automatic repeat request (ARQ) feedback. The authors obtained optimal sensing and access probabilities maximizing the SU throughput under queue stability constraints of both the users and QoS constraint of PU. In \cite{eh_cognitive_4}, authors have considered an interweave scenario where energy harvesting SU tries to maximize it's throughput by accessing the primary channel opportunistically. Authors obtained an optimal transmission policy that maximizes the throughput of the secondary network without colliding with PU's transmission. In \cite{eh_cognitive_5}, authors considered a scenario where an RF energy harvesting SU tries to access the primary channel opportunistically. The primary traffic is modeled as time-homogeneous discrete Markov process. In the beginning of each slot, the SU has to decide whether to remain idle or perform spectrum sensing. Authors jointly designed the spectrum sensing policy and detection threshold to maximize the expected SU throughput under energy causality and collision constraints. In \cite{eh_cognitive_6}, authors proposed a channel selection criterion based on probabilistic availability of energy of SU, PU's belief state and channel conditions and obtained channel aware optimal and myopic sensing policies maximizing throughput of energy harvesting SU in single-user multi-channel interweave scenario. 

In overlay scenario, the SU has knowledge of PU's transmit sequence and encoding scheme. On one hand, this information is used to cancel the primary interference at the secondary receiver and on the other hand, secondary can assist primary in it's transmission by relaying primary data. In \cite{coop_cognitive_shafie}, authors considered a scenario where energy harvesting SU helps primary deliver it's data. The inner and outer bounds on stability region are obtained and impact of energy arrival rate on the stability region is studied. In \cite{yin_cooperation}, authors considered an overlay scenario where energy harvesting SU operates with \emph{save-then-transmit} protocol. The SU can cooperate with PU by relaying it's data. The authors obtained optimal decision (to cooperate or not) and optimal action (how much time for energy harvesting and how much power allocation for relaying) maximizing secondary throughput. Proposed optimal cooperation protocol is then compared with non-cooperation and stochastic-cooperation protocols. In \cite{info_ener_coop_cognitive}, authors have studied joint energy and information cooperation between primary and secondary users. The secondary user acts as a relay for primary user and in return, primary user feeds it with energy. Three schemes enabling this cooperation are studied. In first, an ideal backhaul between the two users is considered. Next, authors introduced power splitting and time splitting techniques for information and energy transfer and obtained zero forcing solution for all the schemes. In \cite{sanket_pimrc}, authors have considered energy and information cooperation between primary (PU) and secondary user (SU) in overlay setting where SU helps PU by relaying it's data and in return, PU allows SU to access it's spectrum. In addition, when SU is energy constrained, PU supplies energy to SU for it's operation. Authors obtained optimal offline power policies for PU and SU which maximizes the primary's rate. The effects of SU rate constraint and finite battery on PU rate and probability of cooperation are also studied.

In underlay cognitive radio, the primary and secondary users coexist in an interference limited scenario. The SU can transmit along with PU as long as secondary interference at primary receiver remains below an acceptable threshold. In \cite{eh_cognitive_2}, authors considered a stochastic geometry model where primary and energy harvesting secondary transmitters are modeled as independent homogeneous Poisson point processes (HPPP). There exist two concentric zones, \emph{guard zone} and \emph{harvesting zone} with primary transmitter (PT) at the center. Secondary transmitter (ST) harvests from PT's transmission if it falls in harvesting zone, transmits if it falls outside the guard zone and remains idle otherwise. Authors obtained maximum spatial throughput of SU under outage constraints for coexisting network. In \cite{rakovic_underlay}, authors have considered a slotted underlay system. In each slot, SU first harvests from primary transmission for some duration and then transmits it's own data taking interference at primary receiver into account. Optimal time sharing between energy harvesting phase and information transfer phase is obtained that maximizes average achievable rate subject to $\varepsilon$-percentile protection criteria for primary system. 

However, to best of our knowledge in underlay cognitive radio networks, energy cooperation among the primary and secondary users has not been considered in the literature. In this paper, we consider a CR network where PU and SU coexist in an underlay scenario. Both of the users have energy harvesting capability and are equipped with finite capacity battery to store the harvested energy. The SU has an energy transfer unit that works independent of data transmission unit. We assume that the channel state information and energy arrival are available non-causally at both the transmitters as in \cite{non_causal_csi}. Under these assumptions, our contributions are summarized as follows:
\begin{itemize}
\item First, we consider a single slot scenario and obtained the optimal transmit power and energy transfer policy that maximizes the number of bits transmitted by ST subject to PU's sum rate constraint. Depending on harvested energy and channel power gains, we analytically obtained the optimal transmit power and transferred energy. 
\item Second, we extend our system model to multi-slot scenario and obtained a suboptimal transmit/transfer policy under offline setting. We have considered different interference scenarios and studied the effect of energy cooperation on the system performance in each case.
\end{itemize}
\indent The rest of the paper is organized as follows. Section II describes the system model, problem formulation is given in section III, the offline transmit power and energy transfer policy is obtained in section IV, results are given in section V and finally, the conclusions are drawn in section VI.

\section{System Model}
\begin{figure}[!t]
\centering
\includegraphics[width=2.8in]{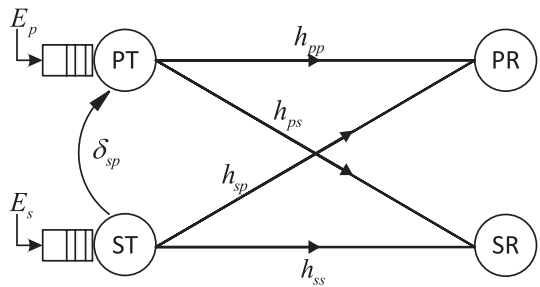}
\caption{Primary and secondary users coexisting in an underlay mode.}
\label{fig:system_model}
\vspace{-0.3cm}
\end{figure}

The system model shown in Fig. \ref{fig:system_model} consists of one energy harvesting pair of primary and secondary transmitter-receiver coexisting in an underlay mode. We assume slotted mode of operation for both the users. At both the transmitters, energy arrives in form of packets at the beginning of each time slot and stored in finite capacity batteries of capacity $E_{max}$. In addition, the secondary transmitter has an energy transfer unit so that it can transfer some portion of it's harvested energy to primary transmitter with efficiency $0\leq \alpha\leq 1$. The energy transfer unit is assumed to be independent of the data transfer unit. All channel links are assumed to be Rayleigh faded and hence, the channel power gains, $h_{pp}^i$, $h_{ps}^i$, $h_{ss}^i$ and  $h_{sp}^i$ are exponentially distributed. Since an offline transmit policy is considered, it is assumed that the harvested energy amounts and arrival times, and channel coefficients are perfectly known to both the transmitters apriori as in \cite{non_causal_csi}.

Both the transmitters transmit for $N$ slots of 1 second each. The PU has total $B_p$ bits to transmit. At the beginning of each slot, ST may transfer some portion of it's harvested energy, $\delta_{sp}^i$ to PT with efficiency $\alpha$ and will use the remaining energy for it's data transmission. In each slot, the transmission powers of both the users are bounded by energy causality constraint, which comes from the fact that total consumed energy upto any slot should always be less than or equal to total energy harvested upto that slot. In addition, finite battery capacity puts an extra constraint on transmission power. Since battery overflow is suboptimal, transmission power needs to be adjusted to avoid battery overflows.
\section{Problem Formulation}
\indent Let $\vec{p}_p$ and $\vec{p}_s$ denote the transmit power vectors of primary and secondary transmitter respectively, $E_p^i$ and $E_s^i$ denote the harvested energy of primary and secondary transmitter respectively in the beginning of $i$th slot. The channel power gains for PT-PR, PT-SR, ST-SR and ST-PR link in the $i$th slot are denoted by $h_{pp}^i$, $h_{ps}^i$, $h_{ss}^i$ and  $h_{sp}^i$ respectively. Under the interference by primary user, the achievable rate of SU in $i$th slot, is given by Shannon's capacity formula: $\log_2\left(1+\frac{h_{ss}^ip_s^i}{\sigma^2+h_{ps}^ip_p^i}\right)$, where $\sigma^2$ is the noise variance. Similarly, the achievable rate of PU is given as: $\log_2\left(1+\frac{h_{pp}^ip_p^i}{\sigma^2+h_{sp}^ip_s^i}\right)$.

Our objective is to find an transmit power and energy transfer policy which maximizes the total number of bits transmitted by secondary transmitter in $N$ slots while allowing the primary user to transmit atleast $B_p$ bits under energy causality and battery overflow constraints for both the transmitters. The optimization problem for no-cooperation case is given as:
\small
\begin{subequations}\label{eq:opt_prob_nocoop}
\begin{align}
\max_{\vec{p}_s,\vec{p}_p} \quad & \sum_{i=1}^N\log_2\left(1+\frac{h_{ss}^ip_s^i}{\sigma^2+h_{ps}^ip_p^i}\right)\label{eq:nocoop_1}\\
& \qquad\qquad\qquad\qquad(\text{SU's sum achievable rate})\nonumber\\
\text{s.t.} \,\,\quad & \sum_{i=1}^N \log_2\left(1+\frac{h_{pp}^ip_p^i}{\sigma^2+h_{sp}^ip_s^i}\right)\geq B_p,\label{eq:nocoop_2}\\
& \quad\quad\quad\quad\quad\quad\quad\quad\text{(PU's sum-rate constraint)}\nonumber\\
&0\leq \sum_{i=1}^jE^i_s- \sum_{i=1}^jp_s^i\leq E_{max}, \quad j=1,\ldots,N,\label{eq:nocoop_3}\\
& \text{(ST's energy causality \& battery overflow constraint)}\nonumber\\
& 0\leq \sum_{i=1}^jE^i_p-\sum_{i=1}^jp_p^i\leq E_{max}, \quad j=1,\ldots,N ,\label{eq:nocoop_4}\\
& \text{(PT's energy causality \& battery overflow constraint)}\nonumber\\
& \vec{p}_s,\vec{p}_p\succeq0.\label{eq:nocoop_5}\\
& \quad\quad\quad\quad\quad\quad\quad\quad\quad (\text{non-zero power constraint}) \nonumber
\end{align}
\end{subequations}
\normalsize
 Similarly, when we consider energy cooperation among the users, the optimization problem can be written as:
\small
\begin{subequations}\label{eq:opt_prob_multi}
\begin{align}
\max_{\vec{p}_s,\vec{p}_p,\vec{\pmb{\delta}}_{sp}} \quad & \sum_{i=1}^N\log_2\left(1+\frac{h_{ss}^ip_s^i}{\sigma^2+h_{ps}^ip_p^i}\right)\label{eq:multislot_1}\\
\text{s.t.}\quad \quad & \sum_{i=1}^N \log_2\left(1+\frac{h_{pp}^ip_p^i}{\sigma^2+h_{sp}^ip_s^i}\right)\geq B_p,\label{eq:multislot_2}\\
&0\leq \sum_{i=1}^jE_s^i-\sum_{i=1}^j\left(p_s^i+\delta_{sp}^i\right)\leq E_{max}, \, j=1,\ldots,N,\label{eq:multislot_3}\\
& 0\leq \sum_{i=1}^j E_p^i-\sum_{i=1}^j\left(p_p^i-\alpha\delta_{sp}^i\right)\leq E_{max}, \, j=1,\ldots,N, \label{eq:multislot_4}\\
& \vec{p}_s,\vec{p}_p,\vec{\pmb{\delta}}_{sp}\succeq0.\label{eq:multislot_5}
\end{align}
\end{subequations}
\normalsize
where $\delta^i_{sp}$ is the energy transferred from SU to PU in the beginning of $i$th slot. The LHS of constraints (\ref{eq:nocoop_3})-(\ref{eq:multislot_3}) and (\ref{eq:nocoop_4})-(\ref{eq:multislot_4}) are the energy causality constraints for secondary and primary transmitter respectively, which represents that the total consumed energy by the end of slot $i$ should always be less than or equal to the harvested energy upto slot $i$. Whereas, RHS of constraints (\ref{eq:nocoop_3})-(\ref{eq:multislot_3}) and (\ref{eq:nocoop_4})-(\ref{eq:multislot_4}) are battery overflow constraints for ST and PT respectively, which represents that the remaining energy in the battery by the end of slot $i$ should be less than or equal to battery capacity.\\

\begin{figure*}[!th]
\small
{{\begin{multline}
\mathcal{L}(\mathcal{X},\mathcal{Y})= -\sum_{i=1}^N\log_2\left(1+\frac{h_{ss}^ip_s^i}{\sigma^2+h_{ps}^ip_p^i}\right)+\mu\left[B_p-\sum_{i=1}^N \log_2\left(1+\frac{h_{pp}^ip_p^i}{\sigma^2+h_{sp}^ip_s^i}\right)\right]
 +\sum_{j=1}^N\lambda_j\left[\sum_{i=1}^j\left(p_s^i+\delta_{sp}^i\right)- \sum_{i=1}^j E_s^i\right]\\
 +\sum_{j=1}^N\nu_j\left[\sum_{i=1}^jE_s^i-E_{max}-\sum_{i=1}^j(p_s^i+\delta_{sp}^i)\right]
+ \sum_{j=1}^N\gamma_j\left[\sum_{i=1}^j(p_p^i-\alpha\delta_{sp}^i)-\sum_{i=1}^jE_p^i\right]+\sum_{j=1}^N\theta_j\left[\sum_{i=1}^jE_p^i-E_{max}-\sum_{i=1}^j(p_p^i-\alpha\delta_{sp}^i)\right]
 \label{eq:lagrangian}
\end{multline}}}
\normalsize
where $\mathcal{X}=\{\vec{p}_s,\vec{p}_p,\vec{\pmb{\delta}}_{sp}\}$ and $\mathcal{Y}=\{\mu,\vec{\pmb{\lambda}},\vec{\pmb{\nu}},\vec{\pmb{\gamma}},\vec{\pmb{\theta}}\}$ are the set of primal and dual variables respectively.\\
\noindent\rule{\linewidth}{0.4pt}
\vspace{-0.6cm}
\end{figure*}
\section{Offline Transmit Policy}
\indent In the case of single-slot scenario, the optimization problem in (\ref{eq:multislot_1})-(\ref{eq:multislot_5}) can be reformulated as \emph{linear fractional program} and solved efficiently using CVX \cite{cvx}. Whereas, for multi-slot scenario, the optimization problem becomes non-convex and a suboptimal solution is obtained using \emph{subgradient} method.
\subsection{Single-Slot Scenario}
For single slot ($N$=1), the optimization problem in \cref{eq:opt_prob_multi} can be written as:
\begin{subequations}\label{single}
\begin{align}
\max_{p_{s},p_{p},\delta_{sp}} \quad & \log_2\left(1+\frac{h_{ss}p_{s}}{\sigma^2+h_{ps}p_{p}}\right)\label{single_1}\\
\text{s.t.} \quad\quad & \log_2\left(1+\frac{h_{pp}p_p}{\sigma^2+h_{sp}p_s}\right)\geq B_p, \label{single_2}\\
& p_{s}+\delta_{sp} \leq E'_s, \label{single_3} \\
& p_{p}-\alpha\delta_{sp}\leq E'_p,\label{single_4}\\
& p_{p},p_{s},\delta_{sp}\geq 0. \label{single_5}
\end{align}
\end{subequations}
where $E'_s=\min\{E_s,E_{max}\}$ and $E'_p=\min\{E_p,E_{max}\}$. Since $\log(1+x)$ is a strictly concave monotonically increasing function of $x$, above problem rewritten as:
\begin{align}\label{single_lfp}
\max_{p_{s},p_{p},\delta_{sp}} \quad & \quad \frac{p_{s}}{\sigma^2+h_{ps}p_{p}}\\
\text{s.t.} \quad\quad & -h_{pp}p_{p}+(2^{B_p}-1)h_{sp}p_{s}\leq -(2^{B_p}-1)\sigma^2, \nonumber\\
& (\ref{single_3}), (\ref{single_4}), (\ref{single_5}). \nonumber
\end{align}
which is a \emph{linear fractional program} \cite{boyd} of the form 
\begin{align*}
\max_{\vec{x}} \quad & \frac{\vec{c}^T\vec{x}+a}{\vec{d}^T\vec{x}+b}\\
\text{s.t.} \quad & \mat{A}\vec{x} \preceq \vec{\pmb{\beta}}.
\end{align*}
where,
\begin{align*}
\vec{c}=\left[
\begin{array}{c}
0\\
1\\
0
\end{array}
\right],
\vec{d}=\left[\begin{array}{c}
h_{ps}\\
0\\
0
\end{array}\right],
\vec{\pmb{\beta}}=\left[
\begin{array}{c}
-\omega\sigma^2\\
E'_p\\
E'_s\\
0\\
0\\
0
\end{array}
\right],
\vec{x}=\left[\begin{array}{c}
p_{p}\\
p_{s}\\
\delta_{sp}
\end{array}\right],
\end{align*}
\begin{align*}
&\mat{A}=\left[
\begin{array}{ccc}
-h_{pp}& \omega h_{sp}& 0\\
1 & 0 & -\alpha\\
0 & 1 & 1\\
-1 & 0 & 0\\
0 & -1 & 0\\
0 & 0 & -1
\end{array}
\right],
a=0 \text{ and } b=\sigma^2.
\end{align*}
where $\omega=(2^{B_p}-1)$.\\
\indent The LFP can be converted into \emph{linear program} (LP) using the change of variables, $\vec{y}=\frac{1}{\vec{d}^T\vec{x}+b}\cdot \vec{x}$ and $t=\frac{1}{\vec{d}^T\vec{x}+b}$ such that $\vec{x}=\frac{1}{t}\cdot\vec{y}$. The equivalent linear program is
\begin{subequations}\label{lp}
\begin{align}
\max_{\vec{y},t} \quad & \vec{c}^T\vec{y}+at \label{lp_1}\\
\text{s.t.} \quad & \mat{A}\vec{y}\preceq t\vec{\pmb{\beta}},\label{lp_2}\\
&\vec{d}^T\vec{y}+bt=1,\, t\geq 0\label{lp_3}
\end{align}
\end{subequations}
which can be solved efficiently using CVX.\\
\textbf{Condition for energy transfer:} Now we find a condition for which energy cooperation occurs and closed form solutions for optimal transmit power and energy transfer.

For no-cooperation case, it is observed that in an optimal policy, constraint (\ref{single_2}) holds with equality since the objective function is monotonically decreasing (increasing) function of $p_p$ ($p_s$) whereas primary's rate is an monotonically increasing (decreasing) function of $p_p$ ($p_s$). Therefore, we have $p_p=\frac{h_{sp}}{h_{pp}}\omega p_s+\frac{1}{h_{pp}}\omega\sigma^2$. Now, we can rewrite the problem as:
\begin{align*}
\max_{p_s}\quad & \frac{p_s}{\sigma^2+\frac{h_{ps}h_{sp}}{h_{pp}}\omega p_s+\frac{h_{ps}}{h_{pp}}\omega\sigma^2}\\
\text{s.t.} \quad & p_s\leq E'_s,\\
& p_s\leq \frac{h_{pp}}{\omega h_{sp}}\left[E'_p-\frac{\omega}{h_{pp}}\sigma^2\right].
\end{align*}
Since, the objective function is a monotonically non-decreasing function of $p_s$, the optimal solution to the problem is given as:
\begin{align}
p_s^{*(nc)}&=\min\left\{\frac{h_{pp}}{\omega h_{sp}}\left[E'_p-\frac{\omega}{h_{pp}}\sigma^2\right],E'_s \right\}\\
p_p^{*(nc)}&=\frac{h_{sp}}{h_{pp}}\omega p_s^*+\frac{1}{h_{pp}}\omega\sigma^2
\end{align}
\indent From the above solutions it is observed that SU transfers its energy if and only if it has non-zero energy left in the battery, that is, there exists a threshold $\zeta= \frac{h_{pp}}{\omega h_{sp}}\left[E'_p-\frac{\omega}{h_{pp}}\sigma^2\right]/E'_s$ such that energy transfer occurs iff $\zeta<1$.\\
\indent In this case, the constraints (\ref{single_2}), (\ref{single_3}) and (\ref{single_4}) will hold with equality and the optimal $p_s,p_p$ and $\delta_{sp}$ are given by the solution of the system of linear equations $\mat{A}\vec{x}=\vec{b}$, where
\begin{align*}
\mat{A}=\left[
\begin{array}{ccc}
-h_{pp} & \omega h_{sp} & 0\\
1 & 0 & -\alpha\\
0 & 1 & 1
\end{array}
\right], \vec{x}=\left[
\begin{array}{c}
p_p\\
p_s\\
\delta_{sp}
\end{array}
\right],
\vec{b}=\left[
\begin{array}{c}
-\omega\sigma^2\\
E'_p\\
E'_s
\end{array}
\right].
\end{align*}
\subsection{Multi-Slot Scenario}
\indent The optimization problem in (\ref{eq:multislot_1})-(\ref{eq:multislot_5}) is a non-convex optimization problem. Therefore, finding an optimal solution to this problem is a difficult task. Hence, we use \emph{subgradient} method to obtain a suboptimal solution of this problem. The Lagrangian of the problem (\ref{eq:multislot_1})-(\ref{eq:multislot_5}) is given in eq. (\ref{eq:lagrangian}) where $\mu,\vec{\pmb{\lambda}}-\vec{\pmb{\nu}}$ and $\vec{\pmb{\gamma}}-\vec{\pmb{\theta}}$ are dual variables for constraints (\ref{eq:multislot_2}), (\ref{eq:multislot_3}) and (\ref{eq:multislot_4}) respectively. In each iteration, the primal and dual variables are updated until convergence in achieved. The primal ($\mathcal{X}$) and dual ($\mathcal{Y}$) variable updates are given as follows:
\begin{align*}
\mathcal{X}^{new}_j&:=\max\{\mathcal{X}_j^{old}-\beta_{\mathcal{X}_j}\cdot(\nabla_{\mathcal{X}_j}\mathcal{L}),0\}\\
\mathcal{Y}^{new}_j&:=\max\{\mathcal{Y}_j^{old}+\beta_{\mathcal{Y}_j}\cdot(\nabla_{\mathcal{Y}_j}\mathcal{L}),0\}
\end{align*}
where $\beta_{\mathcal{X}_j}$ and $\beta_{\mathcal{Y}_j}$ are fixed step sizes for $j$th primal and dual variable respectively. The gradients $\nabla_{\mathcal{X}}\mathcal{L}$ and $\nabla_{\mathcal{Y}}\mathcal{L}$ are given as follows:
\small
\begin{align*}
\frac{\partial\mathcal{L}}{\partial p_p^i}=&\frac{h_{ss}^ih_{ps}^ip_s^i}{(\sigma^2+h_{ps}^ip_p^i+h_{ss}^ip_s^i)(\sigma^2+h_{ps}^ip_p^i)}+\sum_{j=i}^N\gamma_j\\
&\quad\quad\,\,\,-\sum_{j=i}^N\theta_j-\mu\left[\frac{h_{pp}^i}{\sigma^2+h_{sp}^ip_s^i+h_{pp}^ip_p^i}\right]\\
\frac{\partial\mathcal{L}}{\partial p_s^i}=&-\frac{h_{ss}^i}{\sigma^2+h_{ps}^ip_p^i+h_{ss}^ip_s^i}+\sum_{j=i}^N\lambda_j-\sum_{j=i}^N\nu_j\\
&+\mu\left[\frac{h_{pp}^ih_{sp}^ip_p^i}{(\sigma^2+h_{sp}^ip_s^i+h_{pp}^ip_p^i)(\sigma^2+h_{sp}^ip_s^i)}\right]\\
\frac{\partial\mathcal{L}}{\partial\delta_{sp}^i}=&\sum_{j=i}^N\lambda_j+\alpha\sum_{j=i}^N\theta_j-\sum_{j=i}^N\nu_j-\alpha\sum_{j=i}^N\gamma_j\\
\frac{\partial\mathcal{L}}{\partial\lambda_i}=&\sum_{j=1}^i(p_s^j+\delta_{sp}^j)-\sum_{j=1}^iE_s^j, \\
\frac{\partial\mathcal{L}}{\partial\gamma_i}=&\sum_{j=1}^i(p_p^j-\alpha\delta_{sp}^j)-\sum_{j=1}^iE_p^j\\
\frac{\partial\mathcal{L}}{\partial\nu_i}=&\sum_{j=1}^iE_s^i-E_{max}-\sum_{j=1}^i(p_s^j+\delta_{sp}^j)\\
\frac{\partial\mathcal{L}}{\partial\theta_i}=&\sum_{j=1}^iE_p^i-E_{max}-\sum_{j=1}^i(p_s^j-\alpha\delta_{sp}^j)\\
\frac{\partial\mathcal{L}}{\partial\mu}=&B_p-\sum_{i=1}^N\log_2\left(1+\frac{h_{pp}^ip_p^i}{\sigma^2+h_{sp}^ip_s^i}\right)
\end{align*}
\normalsize
\textbf{Stopping criterion:} The algorithm stops if the difference of parameter values in previous and current iteration becomes less than a threshold $\varepsilon$.
\begin{algorithm}[!h]
\caption{Subgradient Algorithm}
\label{algo:suboptimal}
\begin{algorithmic}
\State \textbf{Initialization:} Set $\mathcal{X}:=\{\vec{p}_s,\vec{p}_p,\vec{\pmb{\delta}}_{sp}\}\leftarrow0$, $\mathcal{Y}:=\{\mu,\vec{\pmb{\lambda}},\vec{\pmb{\nu}},\vec{\pmb{\gamma}},\vec{\pmb{\theta}}\}\leftarrow0$ and $iter\leftarrow0$. Initialize $\beta_{\mathcal{X}_j},\beta_{\mathcal{Y}_j}$. 
\Repeat \State{$\mathcal{X}_i^{iter+1}\leftarrow\max\{\mathcal{X}_i^{iter}-\beta_{\mathcal{X}_i}\cdot(\nabla_{\mathcal{X}_i}\mathcal{L})^{iter},0\}$
\State $\mathcal{Y}_i^{iter+1}\leftarrow\max\{\mathcal{Y}_i^{iter}+\beta_{\mathcal{Y}_i}\cdot(\nabla_{\mathcal{Y}_i}\mathcal{L})^{iter},0\}$
\State $iter\leftarrow iter+1$}. 
\Until{$||\mathcal{X}^{iter+1}-\mathcal{X}^{iter}||\leq\varepsilon$.}\\
\Return $\vec{p}_s,\vec{p}_p,\vec{\pmb{\delta}}_{sp}$.
\end{algorithmic}
\end{algorithm}
\section{Results and Discussion}
\begin{figure*}[!t]
\centering
\begin{minipage}{0.3\linewidth}
\centering
  \includegraphics[width=2.3in]{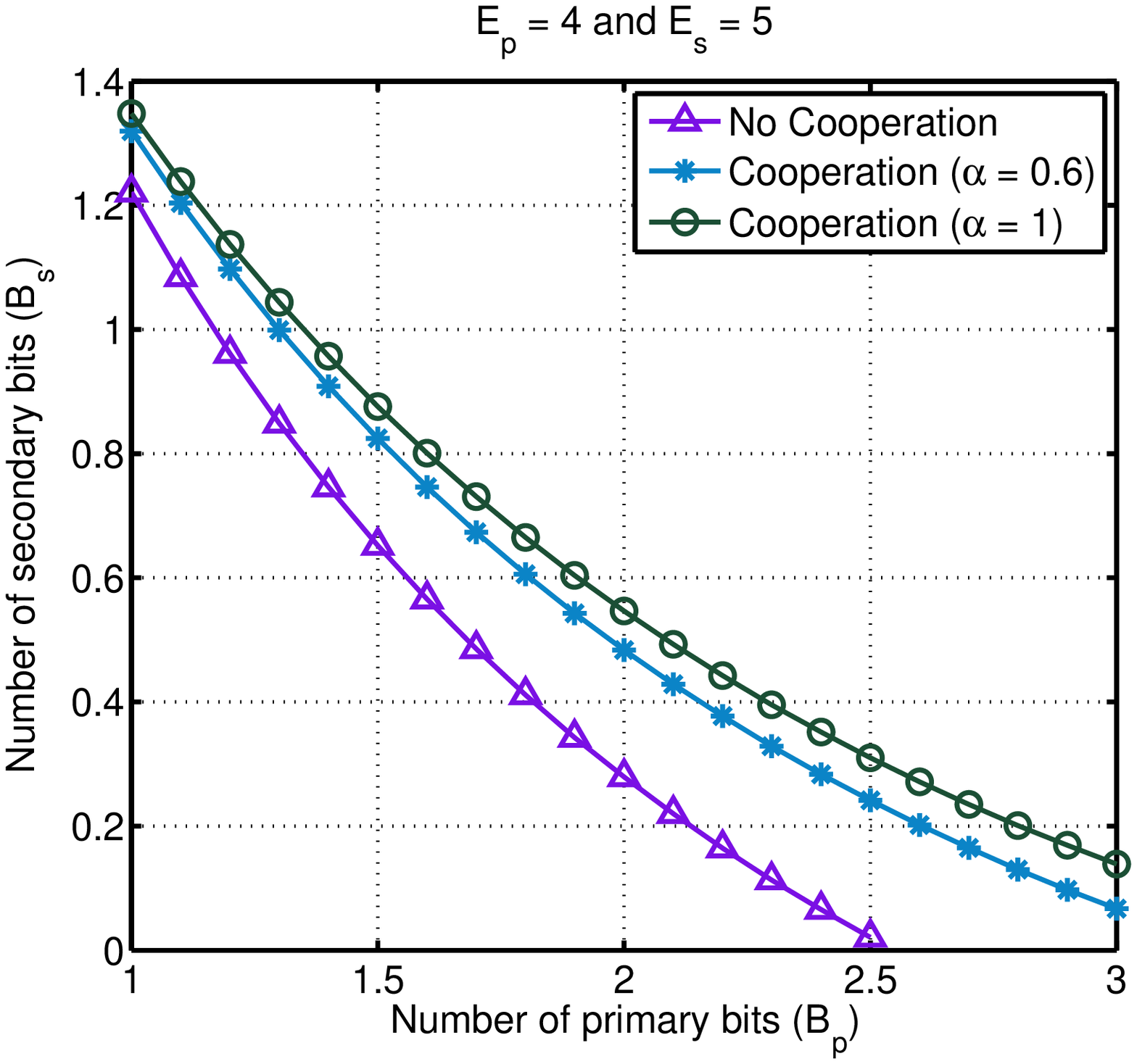}
  \caption{Number of bits transmitted by ST v/s primary sum-rate constraint for single slot scenario as a function of energy transfer efficiency, $\alpha$.}
  \label{fig:single_slot_diff_alpha}
\end{minipage}
\hspace{3mm}
\begin{minipage}{0.3\linewidth}
\centering
  \includegraphics[width=2.3in]{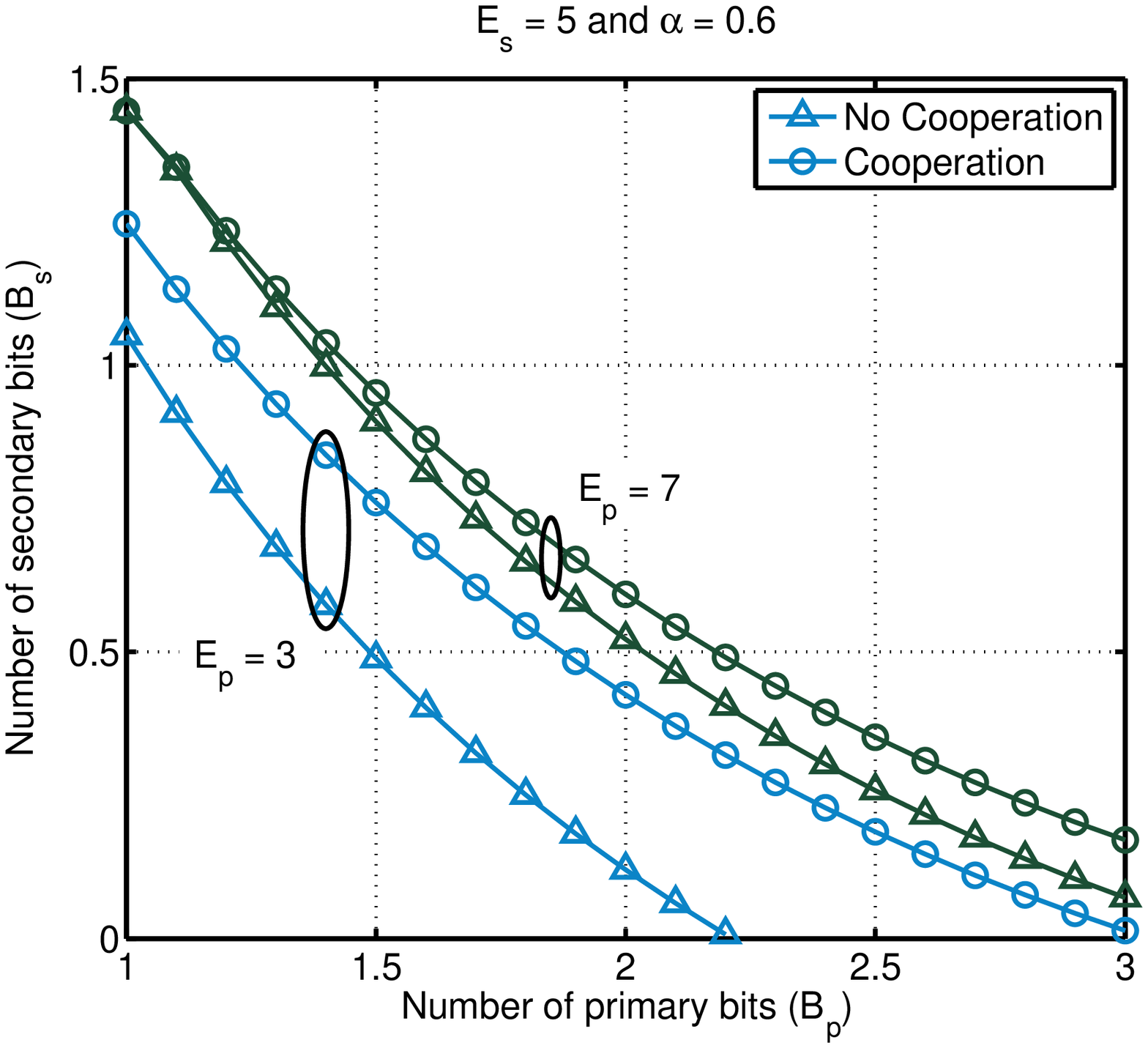}
  \caption{Number of bits transmitted by ST v/s primary sum-rate constraint for single slot scenario as function of energy harvested by PT ($E_p$).}
  \label{fig:single_slot_diff_E_p}
\end{minipage}
\hspace{3mm}
\begin{minipage}{0.3\linewidth}
\centering
\centering
  \includegraphics[width=2.3in]{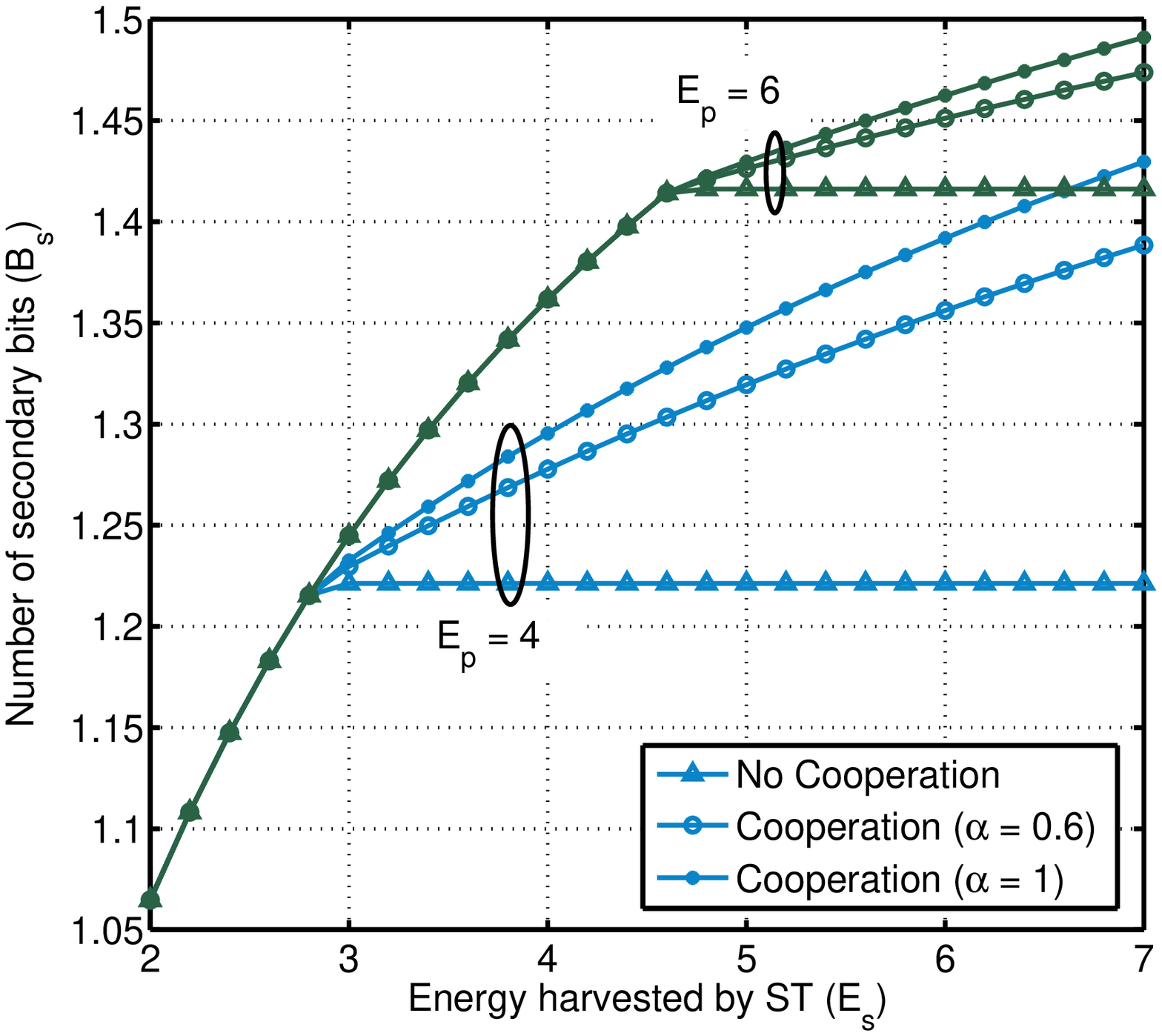}
  \caption{Number of bits transmitted v/s harvested energy of ST for single slot scenario as a function of $\alpha$ and $E_p$ ($B_p=1$).}
  \label{fig:single_slot_B_s_vs_E_s}
\end{minipage}
\vspace{-0.5cm}
\end{figure*}
\subsection{Single-slot scenario}
\begin{figure}[!h]
\centering
 \includegraphics[width=0.85\linewidth]{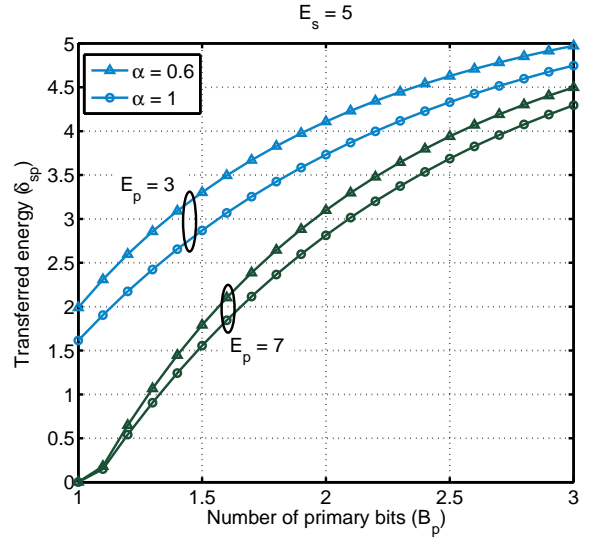}
  \caption{Transferred energy from ST to PT v/s primary sum-rate constraint for single slot scenario as a function of $\alpha$ and $E_p$.}
  \label{fig:single_slot_delta}
  \vspace{-0.5cm}
\end{figure}

\indent The results for single slot case are shown in figures \ref{fig:single_slot_diff_alpha}-\ref{fig:single_slot_delta}. The harvested energy in all the cases is in Joules. All channel links are assumed to be i.i.d. Rayleigh distributed with variances $\sigma^2_{pp}$~=~$\sigma^2_{ps}$~=~$\sigma^2_{sp}$~=~$\sigma^2_{ss}$~=~0.1. 

Fig. \ref{fig:single_slot_diff_alpha} shows the number of bits transmitted by SU as a function of energy transfer efficiency $\alpha$ with respect to PU's sum-rate constraint for single slot scenario. From the figure, it is observed that whenever $\zeta<1$, SU transfers some portion of it's energy to PU. While doing so, it can also increase it's own transmission power so that the constraint (\ref{single_2}) is satisfied with equality (as both PT and ST can now increase their transmission power). This results in higher secondary throughput. Also, it is observed that the number of bits transmitted by ST increases with $\alpha$. This is because as $\alpha$ increases, for same amount of transferred energy, the energy received by PT increases so that ST can now transfer less energy and increase it's own transmission power without violating primary's rate constraint.

Fig. \ref{fig:single_slot_diff_E_p} shows the number of bits transmitted by ST versus primary sum rate constraint for different values of energy harvested by primary user, $E_p$. From the figure, it is observed that as the energy harvested by PT increases, primary can increase it's transmission power which allows the secondary transmitter to increase it's transmission power while maintaining the primary's sum rate constant by keeping the SINR at primary receiver constant (satisfying constraint (\ref{single_2}) with equality). This results in higher secondary throughput. In addition, as $E_p$ increases, the gain by energy transfer reduces as now primary has sufficient available energy to meet it's target rate.

Fig. \ref{fig:single_slot_B_s_vs_E_s} shows the number of bits transmitted by ST versus ST's harvested energy as a function of energy harvested by PT, $E_p$ and energy transfer efficiency, $\alpha$ for fixed primary rate constraint ($B_p=1$). For a fixed $E_p$ and $\alpha$, it is observed that as $E_s$ becomes greater than $\frac{h_{pp}}{\omega h_{sp}}\left[E_p-\frac{\omega}{h_{pp}}\sigma^2\right]$ (i.e., $\zeta<1$), cooperation starts and results in higher secondary throughput. Whereas in no-cooperation case, increased harvested energy does not result in higher throughput because secondary can not increase it's transmission power without violating the rate constraint of PU. For fixed $E_p$, as $\alpha$ increases, the number of bits transmitted by ST increases. The reason of such a behavior is same as given for Fig. \ref{fig:single_slot_diff_alpha}. Also, for fixed $\alpha$, as $E_p$ increases, the threshold $\zeta$ for energy cooperation increases as expected and in addition, the number of transmitted secondary bits increases. The reason for this improvement is same as given for Fig. \ref{fig:single_slot_diff_E_p}.

Fig. \ref{fig:single_slot_delta} shows the transferred energy versus primary rate constraint as a function of $\alpha$ and $E_p$. It is observed from the figure that, for fixed $\alpha$ and $E_p$, as the constraint of PU becomes more strict (i.e. $B_p$ increases), more energy needs to be transferred in order to satisfy the constraint. As $E_p$ increases, since the primary has now sufficient available energy, the energy transferred by ST reduces. Also, it can be observed that as energy transfer efficiency, $\alpha$ increases, ST needs to transfer less energy to PT as for a fixed transferred energy, PT now receives more energy than for the case when $\alpha$ was less. 
\subsection{Multi-slot scenario}
\begin{figure}[!th]
\centering
  \includegraphics[width=.8\linewidth]{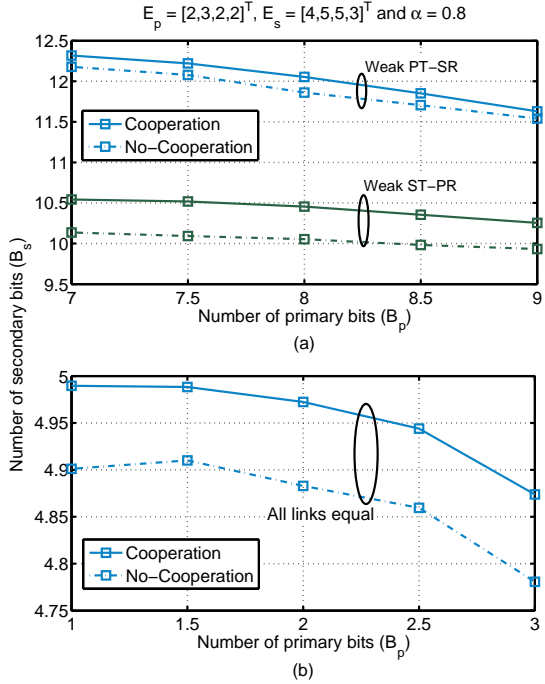}
  \caption{Number of bits transmitted by ST v/s primary sum-rate constraint in multi slot scenario for (a) weak PT-SR and weak ST-PR links (b) equally strong links.}
  \label{fig:multislot_weak_p_s_s_p}
  \vspace{-0.3cm}
\end{figure}

\begin{figure}[!th]
  \centering
  \includegraphics[width=.8\linewidth]{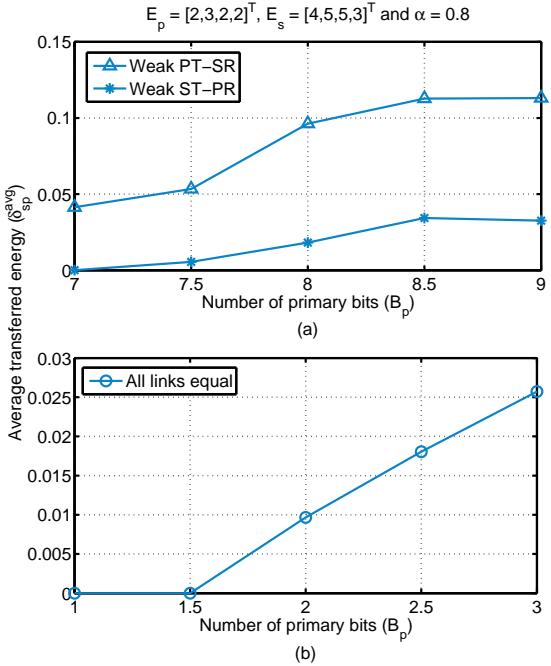}
  \caption{Average transferred energy v/s primary rate constraint in multi slot scenario for (a) weak PT-SR and weak ST-PR links (b) equally strong links.}
  \label{fig:delta_avg_plot_weak_p_s}
  \vspace{-0.3cm}
\end{figure}

\indent The results for the multi-slot case are shown in Fig. \ref{fig:multislot_weak_p_s_s_p} and Fig. \ref{fig:delta_avg_plot_weak_p_s} for $N=4$ under different channel conditions. Under offline setting, the harvested energies for primary and secondary user considered are $E_p=[2,3,2,2]^T$J and $E_s=[4,5,5,3]^T$J respectively. The battery capacity is $E_{max}=6$ J and energy transfer efficiency, $\alpha$ is assumed to be 0.8. All channel links are assumed to be i.i.d. Rayleigh distributed and the noise at both the receivers is assumed to be zero mean white Gaussian with variance $\sigma^2=0.1$.

Fig. \ref{fig:multislot_weak_p_s_s_p}(a) is obtained for weak PT-SR and weak ST-PR links. For weak PT-SR link we have considered $\sigma^2_{pp}=\sigma^2_{ss}=\sigma^2_{sp}=1$ and $\sigma^2_{ps}=0.1$, whereas for weak ST-PR link we have considered $\sigma^2_{pp}=\sigma^2_{ss}=\sigma^2_{ps}=1$ and $\sigma^2_{sp}=0.1$. From the figure, it is clear that when PT-SR or ST-PR links are weak, energy transfer results in better performance. When PT-SR link is weak, better performance is possible because even if ST and PT both increase their transmission power (by allowing energy transfer and maintaining a constant sum rate of primary), due to channel power gains, in SU's sum rate, increment in $p_s$ will dominate the increment in $p_p$ which will result in higher secondary throughput. Similarly, when ST-PR link is weak, higher SU throughput is achieved because the numerator becomes dominant in eq. (\ref{eq:multislot_2}) because of which we can sufficiently increase $p_s$ and simultaneously transfer energy without violating the PU's constraint. Similarly, Fig. \ref{fig:multislot_weak_p_s_s_p}(b) is obtained for equally strong links where we have considered $\sigma^2_{pp}=\sigma^2_{ss}=\sigma^2_{ps}=\sigma^2_{sp}=0.1$. It is clearly observed that energy cooperation results in better performance in terms of number of bits transmitted by ST. This performance improvement can be justified similar to single slot scenario. By transferring some fraction of it's harvested energy, ST can increase it's transmission power so that the number of transmitted bits  increases. And, since the PT has now more energy, it can also increase it's transmission power so that it's transmission rate remains unaffected. Figs. \ref{fig:delta_avg_plot_weak_p_s}(a) and \ref{fig:delta_avg_plot_weak_p_s}(b) show the average $\delta_{sp}$ with respect to primary rate constraint for weak PT-SR and weak ST-PR, and equally strong links respectively. The results show that as the primary rate constraint becomes more strict, ST has to transfer more energy to PT.
\section{Conclusion}
Energy cooperation in underlay cognitive radio network for single as well as multi-slot scenario, is studied and an optimal transmit power and energy transfer policy is obtained for single-slot case, whereas a suboptimal policy is obtained for multi-slot case. From the results, it is observed that in single slot scenario, energy cooperation results in higher secondary throughput. For multi-slot scenario, we obtained the results for weak PT-SR, weak ST-PR and equally strong links and observed that in all cases, energy cooperation improves the performance of secondary system while maintaining the service requirements of primary system. The cases of strong direct links ($\sigma^2_{pp}$~=~$\sigma^2_{ss}$~=~1, $\sigma^2_{sp}$~=~$\sigma^2_{ps}$~=~0.1) and strong interference links ($\sigma^2_{pp}$~=~$\sigma^2_{ss}$~=~0.1, $\sigma^2_{sp}$~=~$\sigma^2_{ps}$~=~1) were also considered but not shown in the paper due to space limitations. In all the cases, energy cooperation results in higher secondary throughput, although in case of strong direct links, gain in secondary throughput due to energy cooperation is not very significant. However, for strong interference links, energy transfer results in significant increase in SU's sum achievable rate. 

%
\bibliographystyle{IEEEtr} 

\bibliography{references}

\end{document}